\documentstyle[epsf,12pt]{article}
\newcommand {\eq}{\begin{equation}}
\newcommand {\qe}{\end{equation}}

\newcommand {\bfr}{{\bf r}}
\newcommand {\bfq}{{\bf q}}

\newcommand {\bfQ}{{\bf Q}}
\newcommand {\ea} {{\it et al.}}

\newcommand {\bfp}{{\bf p}}
\newcommand {\prc}{Phys. Rev. C}

\newcommand {\pr}{{Phys. Rev. }}
\newcommand {\prl}{{Phys. Rev. Lett. }}

\begin{document}
\baselineskip=1.2\baselineskip

\begin{center}
{\large Final State Interactions in Pion Production from Nuclei}

\vspace{.2in}

{\large M. Alqadi and W. R. Gibbs}

\vspace{.15in}

Department of Physics, New Mexico State University, Las Cruces, NM,
88003\\
\today
\end{center}

\vspace{.5in}
  \abstract{We have calculated the effect of the inclusion of final state
interactions on pion production from nuclei with incident pion beams. We 
find that the effect of absorption, along with geometric
considerations, explains much of the enhancement at low invariant mass
seen in recent data for invariant mass spectra with increasing atomic
number. While the variation among nuclei is well reproduced for all
charge states, the enhancement for the ratio for heavy nuclei to
deuterium for the $\pi^+\pi^-$ final state is only partially understood.}

\section{Introduction}

The pion induced pion production on both a free nucleon and a nucleon
within nuclear material provides an opportunity to study the mechanism for
pion production. In a series of papers \cite{chaos96,chaos97,chaos99,
chaos00,chaos01} the results of experiments done at the CHAOS 
(Canadian High Acceptance Orbit Spectrometer) spectrometer
at TRIUMF were presented on pion production with pion beams for several
nuclei.  Specifically, the collaboration reported data for the $\pi^+ A
\rightarrow\pi^+\pi^- X$ and $\pi^+ A \rightarrow\pi^+\pi^+ X$ reactions
on $^{2}$H, $^{12}$C, $^{40}$Ca and $^{208}$Pb at an incident pion kinetic
energy of 283 MeV. The results showed that the shape of the invariant mass
distribution of the two final pions changes with atomic mass number.

An examination of the changes among the heavy nuclei (carbon and heavier) 
reveals that the variation is about the same for the two final charge
states. The shape of the spectrum for the case of a deuteron target is
very different for the two cases. 

A striking feature of the raw spectra is a large peak at small invariant
mass of the two-pion system. This peak is at least partly the result of the
experimental conditions and can be regarded as a Jacobian peak related to
the acceptance of the spectrometer.  Since the correction for this large
instrumental effect is difficult to make, the results are often presented as
ratios, in the expectation that the acceptance effect will cancel. Thus,
Ref. \cite{chaos99} gives ratios to the deuteron cross section, a common way
to normalize. However, in this case, since the deuteron results are very
different for the two final charge states, this procedure results in very
different spectral ratios for them. If one instead takes ratios to the
carbon cross sections, the calcium and lead results are about the same for
the two final charge states.

Shortly after these first data, experiments \cite{cb99} done with the
Crystal Ball at Brookhaven National Laboratory produced data on the
production of two neutral pions for D, C, Al, and Cu with a negative pion
beam.  One sees a very similar trend for the ratios in these data to
that of the CHAOS data.

Since the basic conclusion of these data sets is that the heavier the
nucleus, the more the invariant mass spectrum is peaked toward lower
values of invariant mass a number of groups have seen in this data 
possible evidence for a supposed modification of the $\pi\pi$ interaction
in nuclear matter. 

In early work, a theoretical expectation of an enhancement of strength in
the ${\pi^+\pi^-}$ channel for I=J=0 near the threshold was suggested by
Schuck \ea\cite {schuck88} based on an analogy with Cooper pairs. 

Recently, several studies have placed emphasis on medium effects to
explain the strength enhancement of $M_{\pi\pi}$ distribution in
${\pi^+\pi^-}$ channel ($\sigma$ -meson channel). The calculation of Rapp
\ea \cite{rapp} is based on a simple model in which the pion production
takes place as an elementary reaction. In their model two contributions
were included, a single pion exchange reaction and the excitation followed
by the decay of the ${N^*}$(1440) into two pions and a nucleon. The
$\pi\pi$ final state interaction is included by using the chirally
improved J\"ulich model\cite{julich} and several medium effects were
included. There is reasonable agreement of their calculation and the CHAOS
data in both the ${\pi^+\pi^-}$ and ${\pi^+\pi^+}$ channels.

Vicente-Vacas and Oset used \cite{oset} a macroscopic model for the pion
production with final state interaction among the pions included in the
nuclear medium. Several nuclear effects were taken into account in their
study including an approximate treatment of pion absorption, Pauli
blocking and Fermi motion. Their calculation reproduced the CHAOS data for
calcium in the ${\pi^+\pi^+}$ channel reasonably well and also gave a
reasonable agreement for the deuteron. In contrast, this model failed
to reproduce the CHAOS data in ${\pi^+\pi^-}$ channel except for $^2H$. 

The enhancement of strength in the I=J=0 channel $\pi^+\pi^-$ channel
near the threshold was argued by Hatsuda et al.\cite {hatsuda} to be due
to the partial restoration of chiral symmetry in nuclear matter. 

Davesne \ea \cite{davesne} included the effect of chiral symmetry
restoration and the influence of collective nuclear pionic modes and
found an enhancement of the spectral function a small energy.

Aouissat \ea\cite{Aouissat} argued that the combination of effects of the
restoration of chiral symmetry \cite{hatsuda} in nuclear matter and standard
many body correlation could explain the strength of the spectrum 
$M_{\pi\pi}$ near threshold\cite{schuck88}.

While these considerations may be correct there may also be more prosaic
reasons for the increasing strength at low invariant mass.  We
investigate the effect of the final state interaction of the two
individual pions with the nucleus, especially the absorption of the pion.
Our calculation contains no free parameters and pays special attention to
the geometry of the reaction.

We make two simplifying assumptions.

1) We take the reaction to be coherent on the nucleus.  That is, the
nucleus is assumed to carry off the remaining energy and momentum as a
single entity.  Since it is much heavier than the pions, it can carry a
large amount of momentum with negligible amount of energy.  In this
extreme case the total beam energy is passed on to the two-pion system. 
The invariant mass of the two-pion system is given by

\eq
M=2\sqrt{\mu^2+q^2}
\qe
where $\mu$ is the pion mass (charged or neutral as appropriate) and 
$\bfq$ is the momentum of one of the pions in the two-pion rest frame,
the internal momentum.  The above assumption then leads to the relation
\eq
\omega=\sqrt{M^2+Q^2}
\qe
where $\omega$ is the initial beam energy and $\bfQ$ is the total momentum
of the two-pion pair.

2) We assume that the production takes place at a fixed point on the
surface of the nucleus in a plane at the equator perpendicular to the beam
direction.  The reaction probability is greatly enhanced in this region
for two reasons. First, there is a simple geometrical factor of the
distance to the center of the nucleus. Second, if the incident pion comes
in with a small impact parameter either the initial pion will be scattered
(even a moderate loss of energy will make the pion production
impossible) or one of the final pions will very likely be absorbed, having
to pass through the entire nucleus to escape. Thus, the reactions in which
the incident pion has a small impact parameter are greatly suppressed. 
This point is discussed in Ref. \cite{oset}

\section{Technique}

\subsection{Qualitative Overview}

It is possible to see in a simple qualitative manner that pion absorption
in the final state will lead to a relative enhancement of the invariant
mass spectrum at low pion invariant masses.

\begin{figure}[p]
\epsfysize=190mm
\epsffile{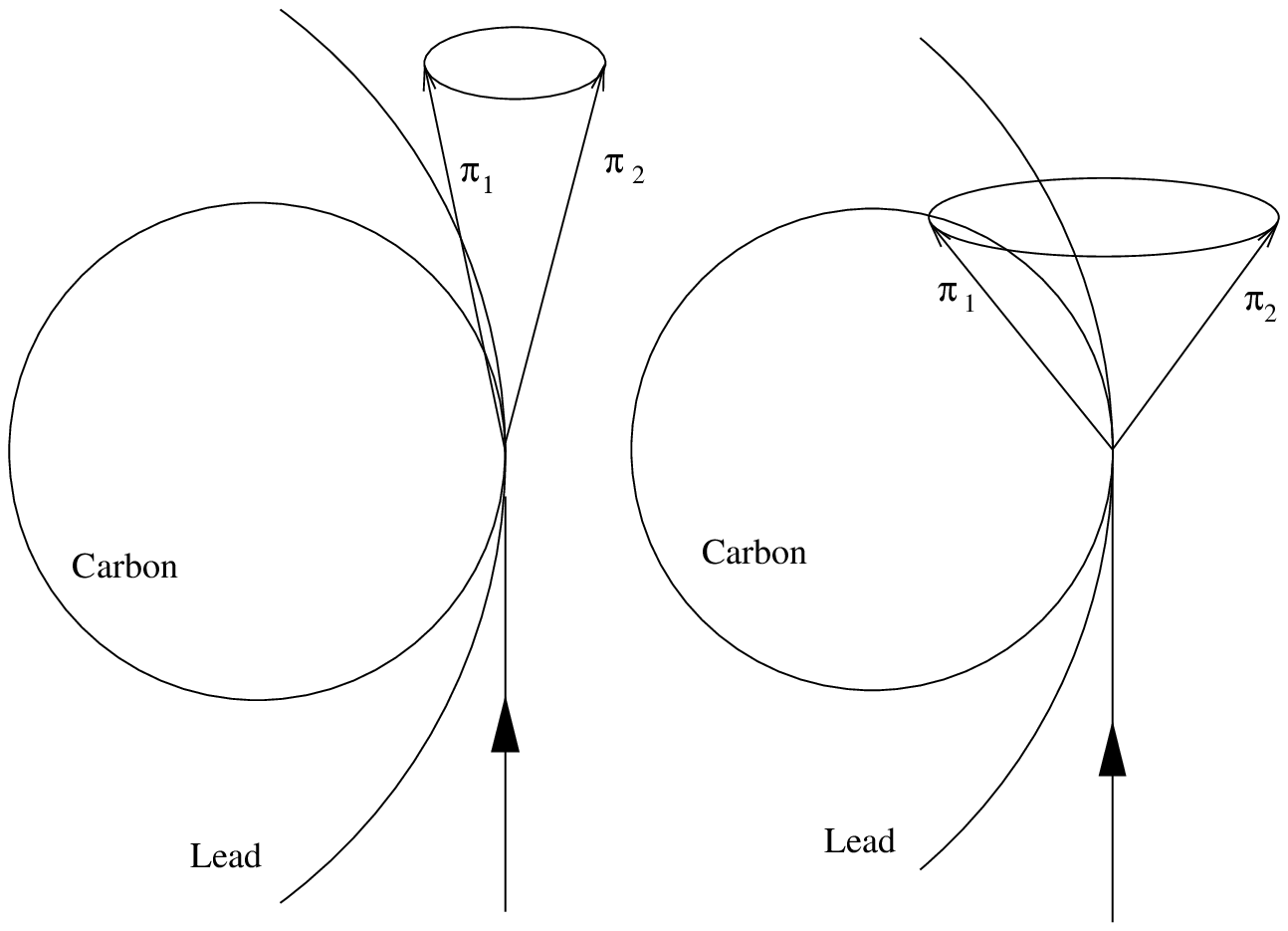}
\caption{Schematic view of the basic effect.}
\label{cone}
\end{figure}

For a first orientation we will assume that the pion pair proceeds forward
in the direction of the initial beam since the distribution of the
momentum of the pair is expected to be very forward peaked.  For the
purposes of the diagram only (see Fig. \ref{cone}) we also take the
momentum of one pion in the frame of the two pion system, $\bfq$, to be
perpendicular to the beam direction. 

With these simplifying assumptions we may now consider the effect of the
absorption of the pions in the final state on the shape of the
invariant mass spectrum.  Figure \ref{cone} shows two cases. The left
portion of the sketch corresponds to the situation where 
the invariant mass is small, so the internal momentum is small and the
momentum of the total two-pion system is large, leading to a narrow cone
representing the paths of the two pions.  It is seen that the pions are
exposed to relatively little interaction with nuclear matter.  In the
limit of the internal momentum becoming zero, the pions would encounter no
higher nuclear density than that in which they were formed.  In this limit
the size of the nucleus would matter very little. 
  
In the right portion of the diagram a case is illustrated in which the
invariant mass and internal momentum are larger.  In this case it is seen
that a large fraction of the pions must traverse a significant amount of
nuclear matter. The larger the nucleus, the larger the probability that
one member of the pion pair would get absorbed. 

Hence, we can anticipate that the geometry of the absorption in the final
state leads to a depletion of the higher values of the invariant-mass
spectrum and that the effect becomes more important with increasing A.

\subsection{Calculation}

The method used here is very similar to that of Oset and Vicente Vacas
 \cite{ov}.  However, they calculated the effect only for calcium and used
an eikonal approximation for the propagation of the final pions. 

We assume that the pion pair is formed quickly at short range by the
strong interaction, and that this formation is independent of the nucleus
that it is near.  Also, we assume that the final total momentum of the two
pions is in the same direction as the incident beam.  This assumption of
exact forward propagation of the pair was investigated by calculating with
finite angles. It was found that the dependence on this angle was very
small.  We define the spherical angles $\theta$ and $\phi$ from the beam
axis as the direction of the $\pi^+$ in the center of the mass of the
pair.

After formation the pions are allowed to propagate independently in 
Coulomb and strong nuclear potentials with relativistic motion.

Their momenta and coordinates  are propagated with the equations 
\eq
 \bfp(t+\delta t)= \bfp(t)+{\bf F} \delta t 
\qe
and 
\eq
 \bfr(t+\delta t)=\bfr(t)+\frac{\bfp(t)}{E(t)} \delta t
\qe
where $\bf F$ is the resultant of the Coulomb and strong interaction
forces. We take the electric charge distribution of the nucleus to be 

\eq
 \rho_c(r)=\frac{\rho_0}{1+e^{\frac{r-\gamma}{\alpha}}}. 
\qe  
The values of $\gamma$ and $\alpha$ were taken from Ref. \cite{devries}
and are (in fm) (2.44, 0.5), (3.07, 0.519), (3.65, 0.54), (4.252, 0.589) and
(6.6, 0.55) for C,  Al, Ca, Cu and Pb respectively. The constant $\rho_0$
was fixed by normalizing the integral of the density to the total charge Ze. 

The nuclear potential is given by \eq V_{s}(r)=-\frac{2\pi}{mA}\left(N
f_{\pi n}+Zf_{\pi p}\right)\rho(r)  \qe where N, Z are the number of
neutrons and protons and A is the atomic mass number. $\rho(r)$ is the
nuclear strong density, i.e. the density of the centers of the nucleons.
\eq 
\rho(r)=\frac{\rho_0}{1+e^{\frac{r-c}{a}}} 
\qe 
The values of c and a were (2.25, 0.5), (2.95, 0.5), (3.55, 0.54), (4.23,
0.55) and (6.5, 0.55) for C, Al, Ca, Cu and Pb respectively and $f_{\pi
n}$ and $f_{\pi p}$ are the real parts of the forward amplitudes of the
${\pi n}$ and ${\pi p}$ amplitudes\cite{gwu}.  This real potential is
modeled after the pion-nucleus optical potential. We will see later that
these non-absorptive strong interactions play a minor role in the result. 

For each initial value of the invariant mass ($M_i$) we calculate
the internal momentum, the total momentum of the two-pion pair and,
with values of the angles of these two quantities, the initial values
of the momenta for each pion can be determined. The pions are then
propagated to a large distance (into free space) and a new invariant
mass is calculated for the pair ($M_f$). In this way the function
\eq
M_f\left(x,\phi,M_i\right),
\qe
where $x=\cos\theta$, is established. It depends on the angles of both
$\bfq$ and $\bfQ$ but we suppress the dependence on the angles of $\bfQ$.

The final spectrum (without absorption) will be given by 

\eq
S_f(x,\phi,M_f)=\frac{d\sigma}{dM_f}=
\frac{d\sigma}{dM_i}/\frac{dM_f}{dM_i}= S_i(x,\phi,M_i)/\frac{dM_f}{dM_i}. 
\qe 
Because the derivative in the denominator is sometimes zero, sharp peaks
appear in the spectra for individual values of the momentum directions. 
The integration over the angles results in a smooth spectrum but, since
the integration is done numerically, some care is required to obtain a
presentable curve.  The lines shown in the figures have been smoothed
since the final calculated values still show a small residual ripple. 

We use as a simple model for the initial spectrum two body phase space.

\subsection{Absorption Factor}

\begin{figure}[p]
\epsfysize=180mm
\epsffile{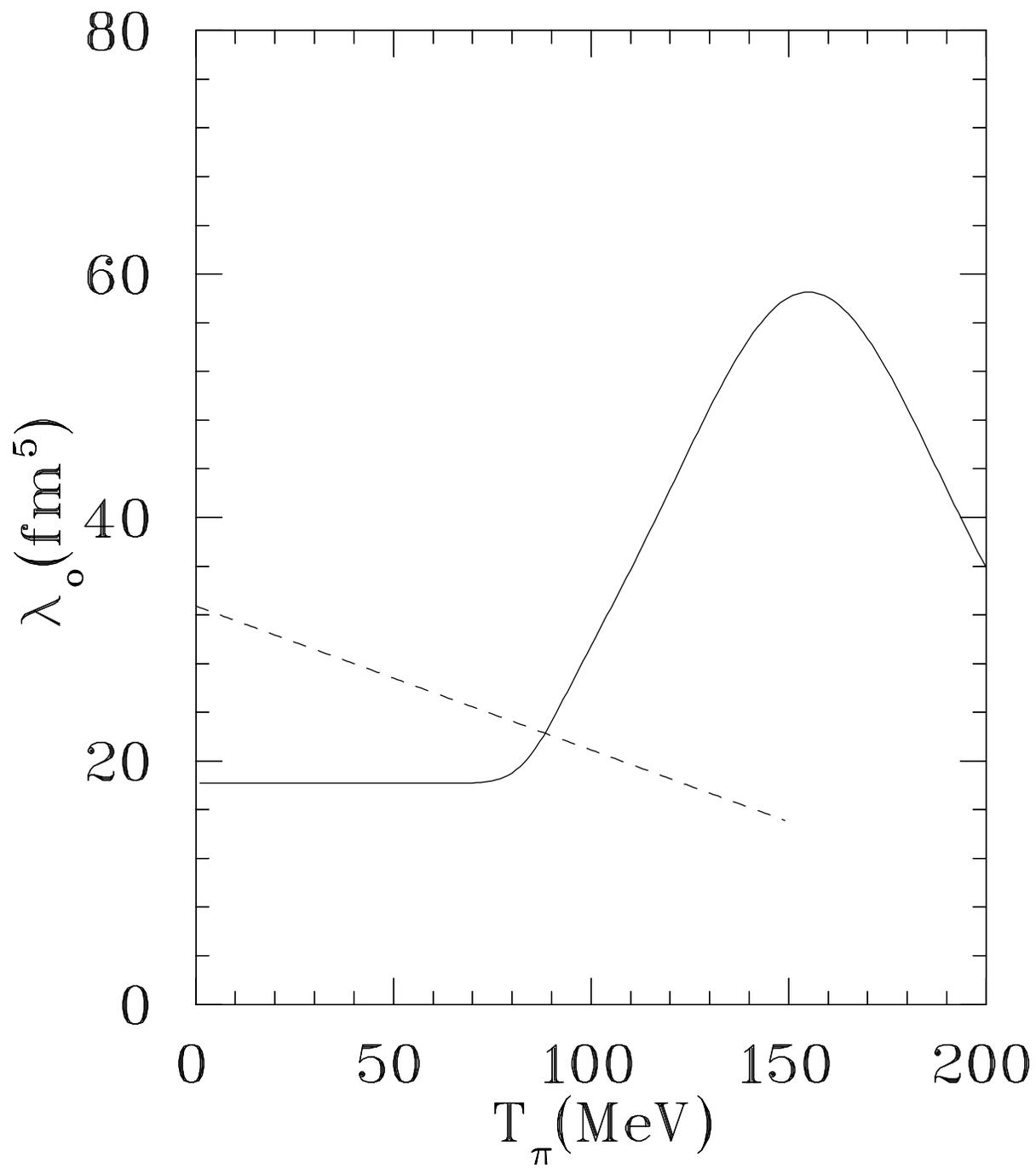}
\caption{$\lambda_0$ determined for light nuclei (solid curve)
and for gold (dashed curve).}
\label{lambda}
\end{figure}

Pion absorption is an important factor during the passage of pions through
the residual nucleus. It reduces the cross section for pion production and
affects the shape of spectrum.  We include the effects of absorption by
calculating the factor the probability that the pion has not been
absorbed, P($\bfr$), at the point $\bfr$ as follows.  The probability of
survival is given by :

\eq
P(\bfr)=1-\int^{\bfr}_{\bfr_0} \lambda(\bfr') P(\bfr')d{{\bf l}} 
\qe
where  $\bfr_0$  is the formation point, and the
initial value of $P\equiv 1$. The integral is taken along the trajectory
of the particle. As in Ref. \cite{ov}, the absorption probability per unit
length is assumed to be proportional to the square of the nuclear density

\eq
 \lambda(r)=\lambda_{0}(E)\rho^2(r),
\qe
The  $\lambda_0 $ parameter depends on the energy of the pion and is
obtained by fitting to absorption data. For light nuclei we first
calculated the eikonal expression for the total absorption cross section
\eq
\sigma(\lambda_0)=2\pi \int^{\infty}_{0} a da \left(1 -
e^{-\int_{-\infty}^{\infty} dz \lambda_0  \rho^{2} (\sqrt{z^2+a^2})}
\label{eik}\right). 
\qe
By comparing the values of $\sigma(\lambda_0)$ with experimental
values of $\sigma_{abs}(E)$ \cite{ashery,nakai,saluk,balan,laber,kirz} we
inferred the values of $\lambda_0(E)$ needed to fit the data (see figure
\ref{lambda}).  We found that the values of $\lambda_0(E)$ had little
dependence on the atomic mass number A for light nuclei. 

\begin{figure}[p]
\epsfysize=180mm
\epsffile{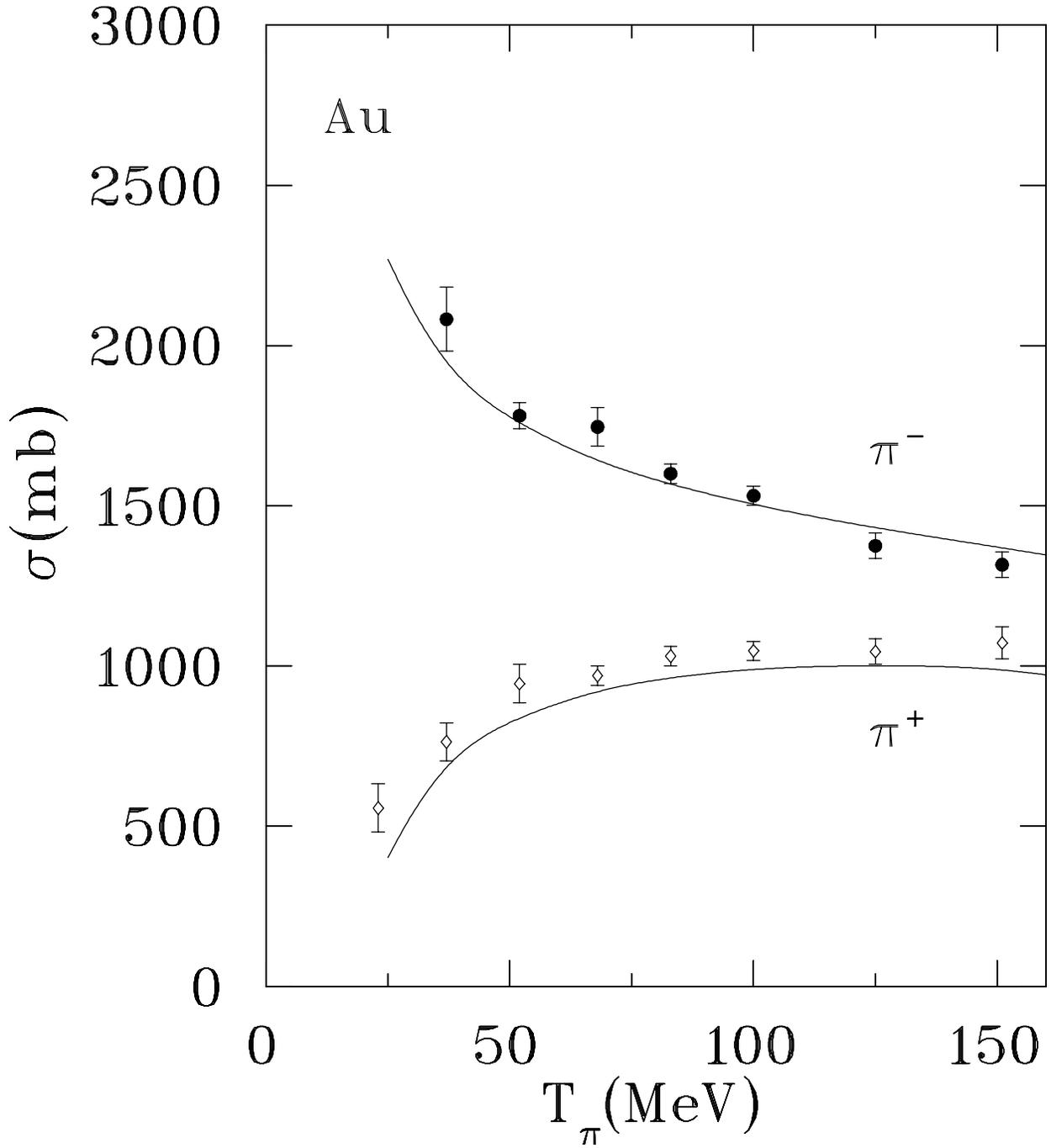}
\caption{Comparison of the absorption cross section computed with the
dashed curve in figure \protect\ref{lambda} with the absorption data
of Nakai \ea \cite{nakai}.}
\label{nakai}
\end{figure}

For heavy nuclei it is important to follow the trajectory instead of using
Eq. \ref{eik}. Figure \ref{nakai} shows the result of the fit for gold.
The $\lambda_0$ used in this case is also given in Fig. \ref{lambda} 

Denoting the result of following the trajectory of the i$^{th}$ pion into
free space by $P_i(x,\phi,M_i)$ the final spectrum is given by the
integral over all possible decay angles,
\eq
S_f(M_f) = \frac{1}{4\pi} \int^{1}_{-1} dx \int^{2\pi}_{0} d\phi
S_f(x,\phi,M_f)P_1(x,\phi,M_i)P_2(x,\phi,M_i).
\qe
        
\section{Results}

\begin{figure}[p]
\epsfysize=180mm
\epsffile{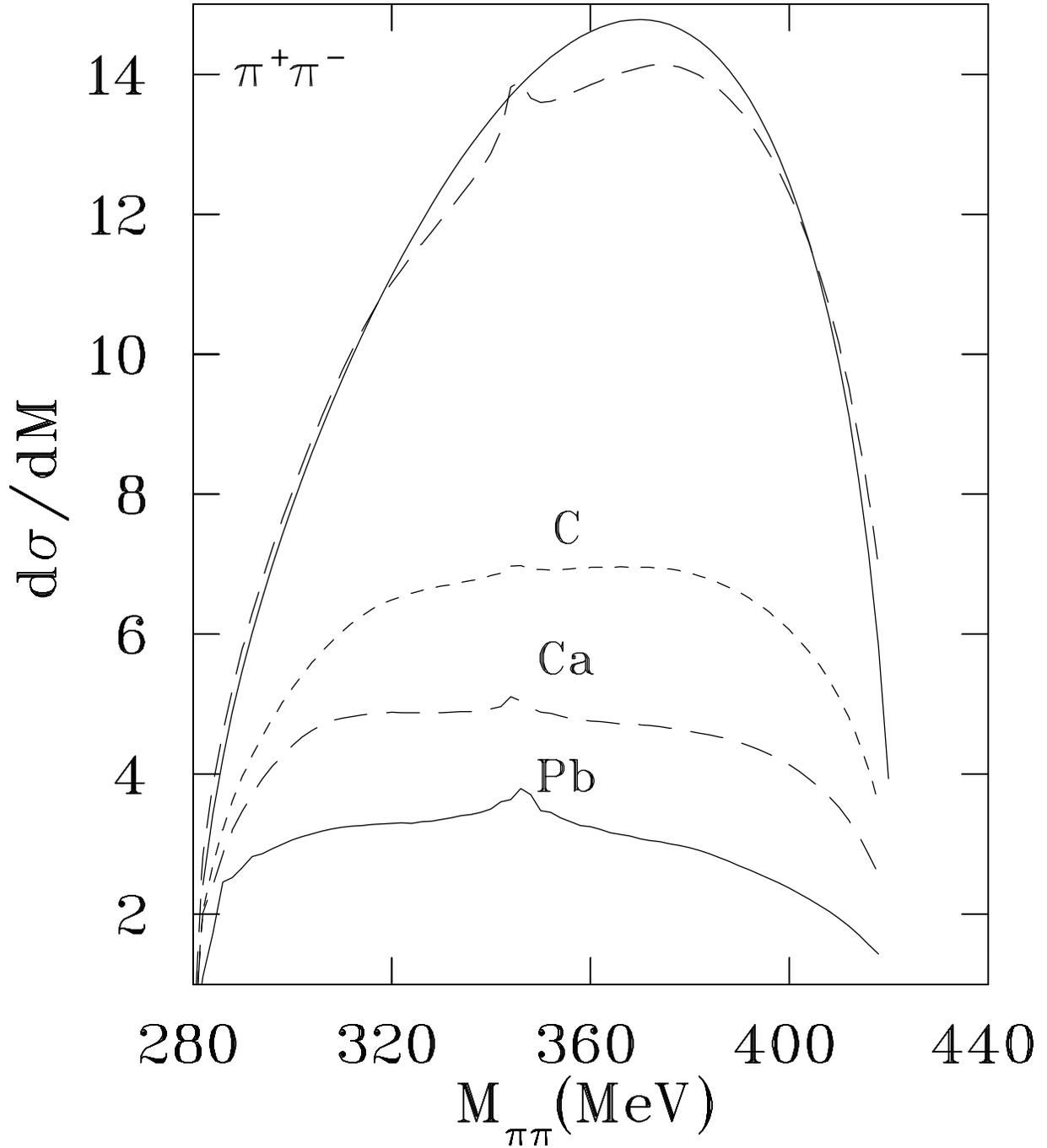}
\caption{Variation of the basic spectrum for carbon, calcium and lead.
The solid curve shows the shape the initial free spectrum and the
long-dashed curve represents the modification due to the non-absorptive
interactions only.}
\label{spectfall}
\end{figure}

Figure \ref{spectfall} shows the resulting spectra as well as the original
model spectrum (solid line). Also shown is the spectrum with only the
strong and Coulomb potentials. These spectra cannot easily be compared
directly to the data for two reasons. \\
1) Our initial model spectrum represents only roughly the free spectrum
for $\pi^+\pi^+$ and not well at all for $\pi^+\pi^-$. In fact, the free
spectrum is not well known. Even if it were, we cannot be
sure that it is the appropriate starting spectrum in the nuclear case. \\
2) There is a large instrumental acceptance correction to be made.

For these reasons we, like the experimental group, compare with ratios.
Since we are interested in the increase of the spectrum ratios near 
small values of invariant mass we normalize the ratios (theoretical and
experimental) to unity at 330 MeV.

The CHAOS group calculated the  ratio
$\frac{^AS_{\pi\pi}}{^DS_{\pi\pi}}$ for both $ \pi^+\pi^- $ and
$\pi^+\pi^+$ to show the effect of the nucleus. The strong
enhancement near threshold is due to the fact that the invariant mass
distribution of D ($^2$H)  and A ($^{12}$C, $^{40}$Ca,$^{208}$Pb) change
dramatically near the threshold where is a much smaller peak in the D
distribution. 

In our study we calculated the ratio of heavier nuclei to carbon instead
of deuterium and found that the ratio increases with A for all
charge channels by almost the same amount.

\begin{figure}[p]
\epsfysize=180mm
\epsffile{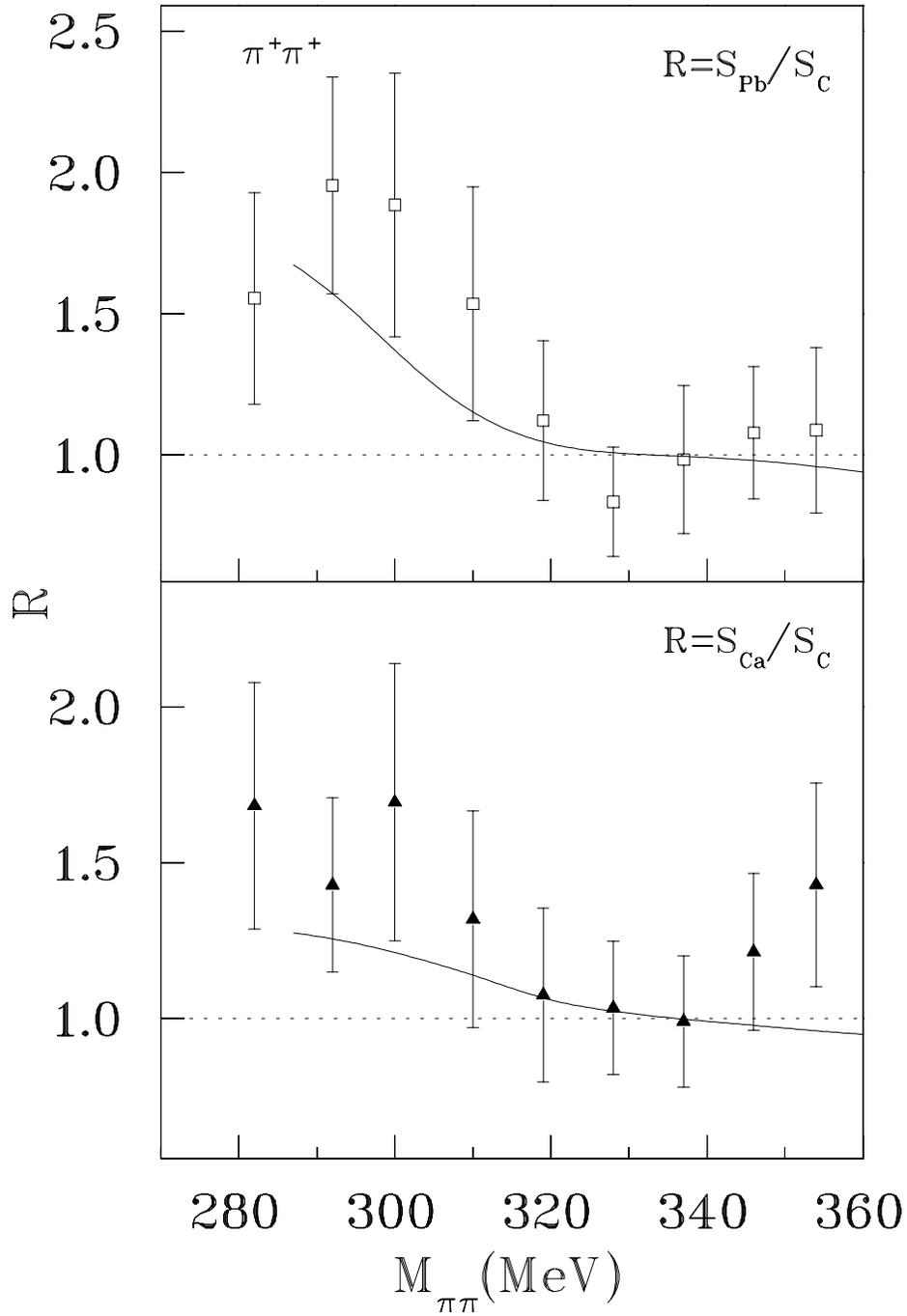}
\caption{Ratios for lead and calcium to carbon for the $\pi^+\pi^+$ final
state compared with the CHAOS data \protect \cite{chaos99}.}
\label{pifpp}
\end{figure}

Figures \ref{pifpp} and \ref{pifpm} show the ratios of the cross sections
for $^{208}$Pb and $^{40}$Ca to that of $^{12}$C. The calculated ratios
are seen to reproduce the experimental data in the $\pi^+\pi^+$ and
$\pi^+\pi^-$ channels.  The two charge channels show similar
behavior. Figure \ref{pifpm} also shows the effect of varying the assumed
radius at which the production takes place.

The upper part of Fig. \ref{lamden} shows the difference seen in the
lead to carbon ratio using the value of $\lambda_0$ determined
from the gold data and that obtained from the light element fit.
The lower part of the same figure shows the difference between 
using the Woods-Saxon density and a shell model density for calcium 
\cite{gd}.

\begin{figure}[p]
\epsfysize=180mm
\epsffile{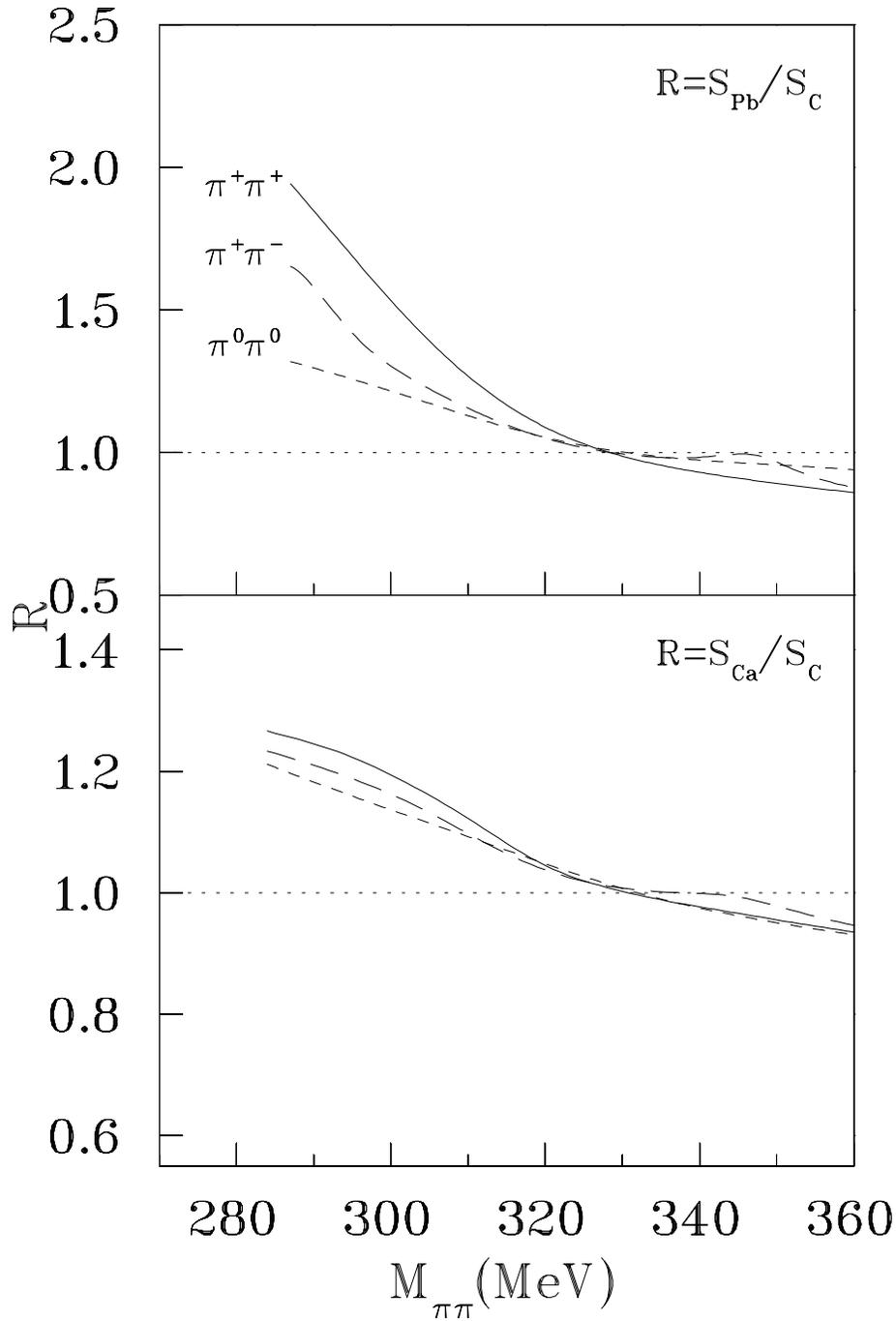}
\caption{Comparison of results for different charge states for the
lead to carbon ratio. The solid curve corresponds to the $\pi^+\pi^+$
final state the short dash to the neutral final state and the long dash to
the $\pi^+\pi^-$ final state.}
\label{charge}
\end{figure}

Figure \ref{charge} shows the dependence on the final
charge state of the calculations. It is seen that the effect is moderate.

\begin{figure}[p]
\epsfysize=180mm
\epsffile{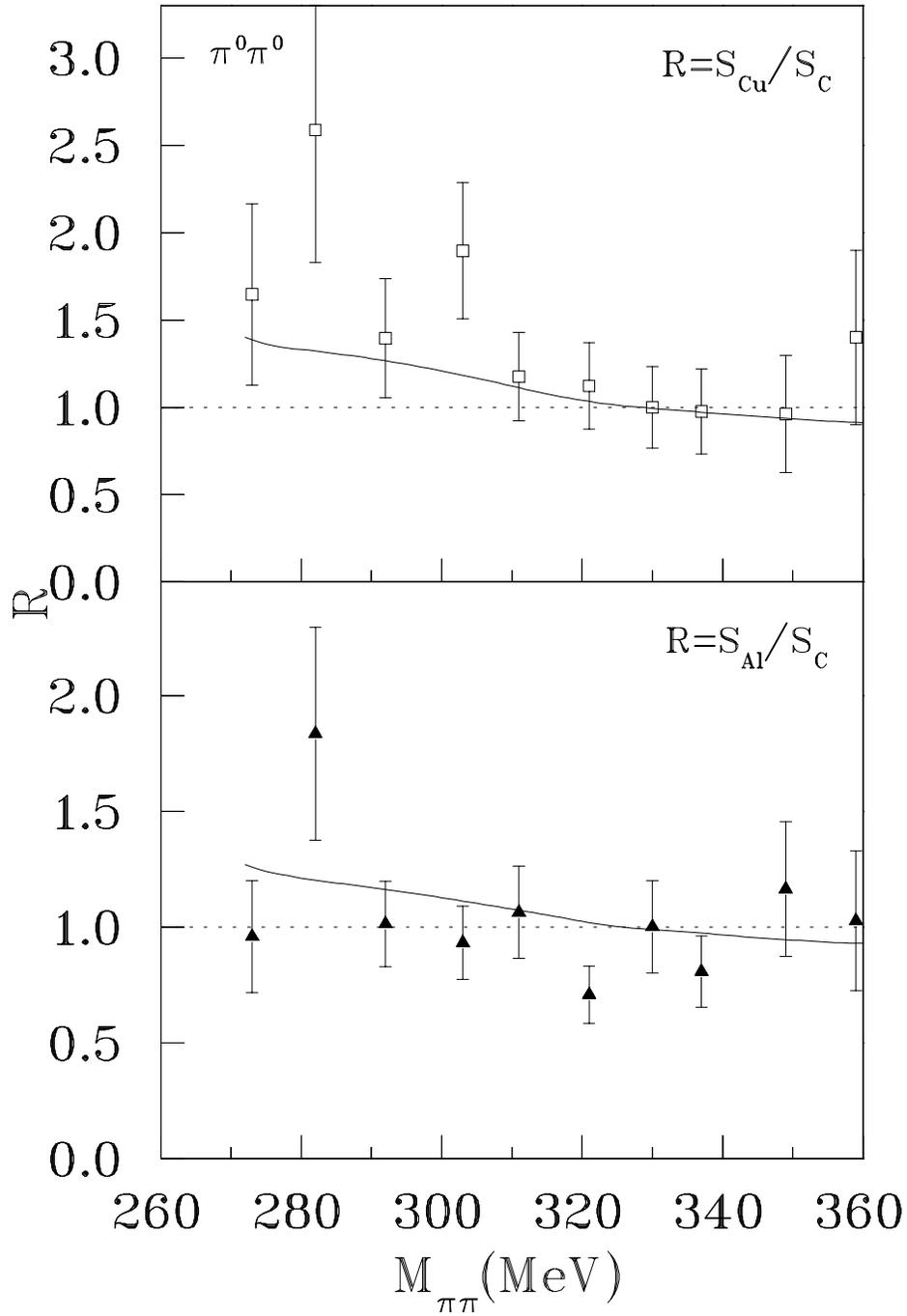}
\caption{Results for the ratios for the $\pi^0\pi^0$ final state compared
with the Crystal Ball data \protect \cite{cb99}.}
\label{pif00}
\end{figure}

Figure \ref{pif00} shows the ratio of the cross sections for Cu and Al to 
C compared to the experimental data in the $\pi^0\pi^0$ channel. Although 
the error bars are large, the trend of the data is clearly reproduced.

The estimate of the interaction potential to use in a classical
calculation by a quantum model must, of course, be approximate.
Figure \ref{pifvs0} shows that this potential plays a minor role by
comparing with calculations with it set to zero.

\begin{figure}[p]
\epsfysize=180mm
\epsffile{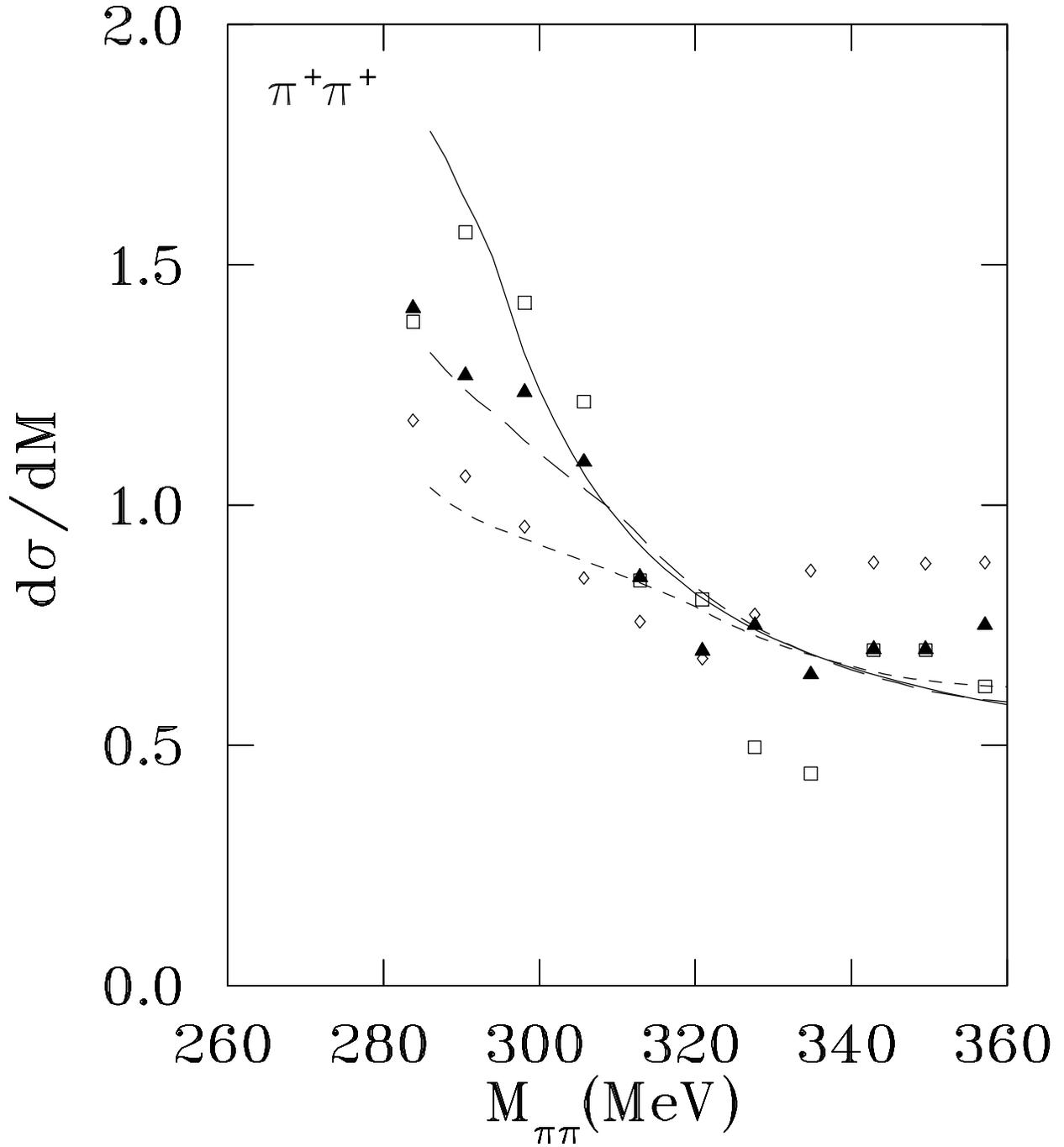}
\caption{Ratios of the spectra of heavy nuclei divided by the basic
spectrum for the $\pi^+\pi^+$ final state. The squares and solid curve
correspond to lead, the triangles and long-dashed curve to calcium
and the diamonds and short-dashed curve to carbon.}
\label{spepp}
\end{figure}

Figure \ref{spepp} shows the ratios of the heavy elements to
deuterium for the $\pi^+\pi^+$ final state. The comparison
between theory and experiment is satisfactory.

\begin{figure}[p]
\epsfysize=180mm
\epsffile{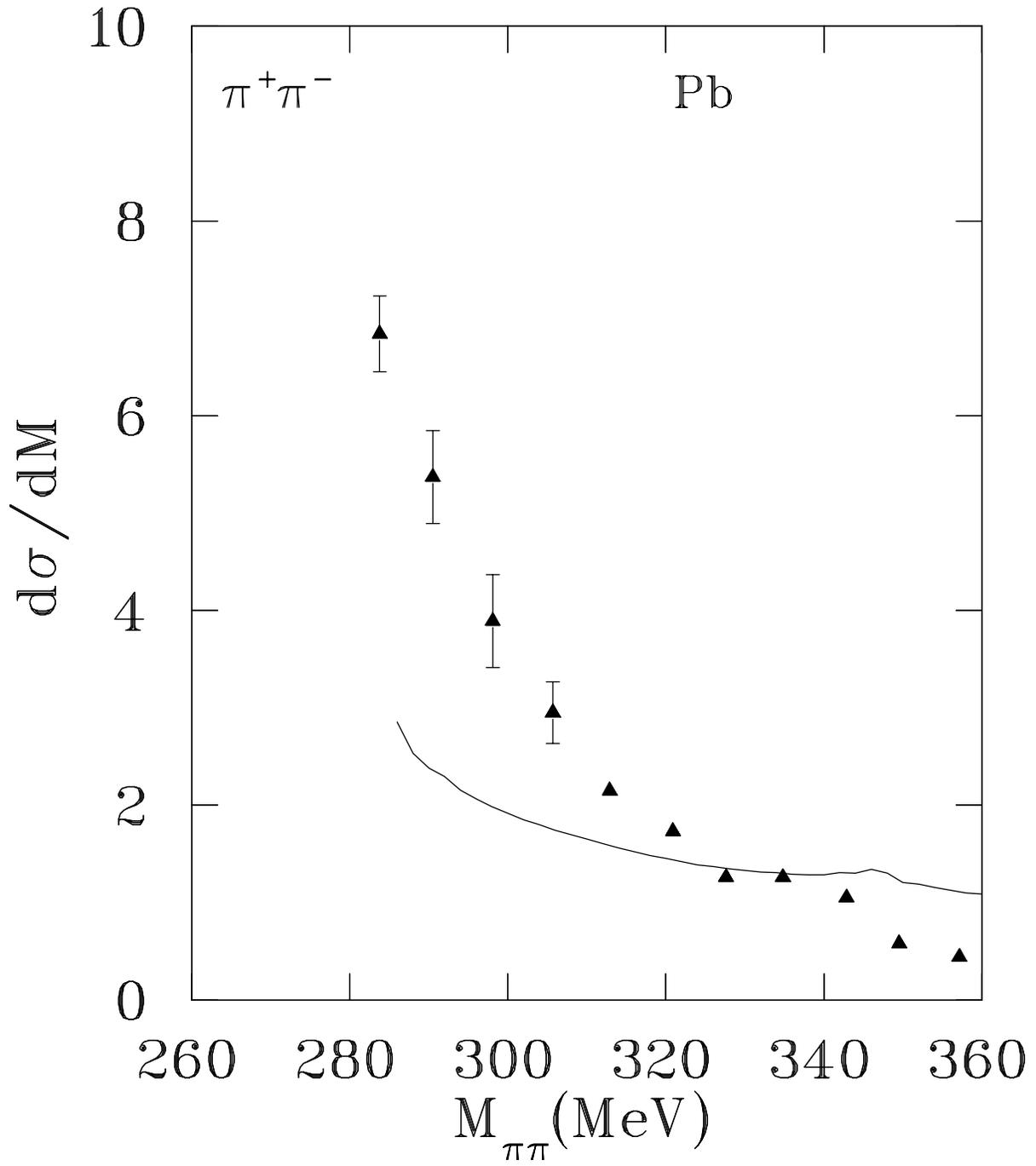}
\caption{Ratio of the lead spectrum to the basic spectrum for the
$\pi^+\pi^-$ final state.}
\label{spectpb}
\end{figure}

Figure \ref{spectpb} shows the ratio of the lead spectrum to
that of deuterium for the $\pi^+\pi^-$ final state. The effect considered
here is able to explain only a little less than half of the observed
experimental effect.

\section{Discussion}

We have seen that the variation in the spectral shape from nucleus to
nucleus (above deuterium) can be understood in terms of the interaction of
the produced pions with the nucleus.  The size of the
nucleus plays a strong role in this explanation.  Thus, there is little or
no evidence in these ratios for a modification of the pion production or
pion-pion interaction.  It is not clear that one should expect an effect of
this type in any case.  As has been discussed, the point at which the
production takes place is largely determined by the density. Thus, as one
moves to heavier nuclei there will be a strong tendency for the reaction to
occur at the same density. Since the average density is the major parameter
used to determine the medium effect, and it tends to remain constant, one
might believe that no nuclear matter effect on the production would be seen
in comparing among heavy nuclei. 

\begin{figure}[p]
\epsfysize=180mm
\epsffile{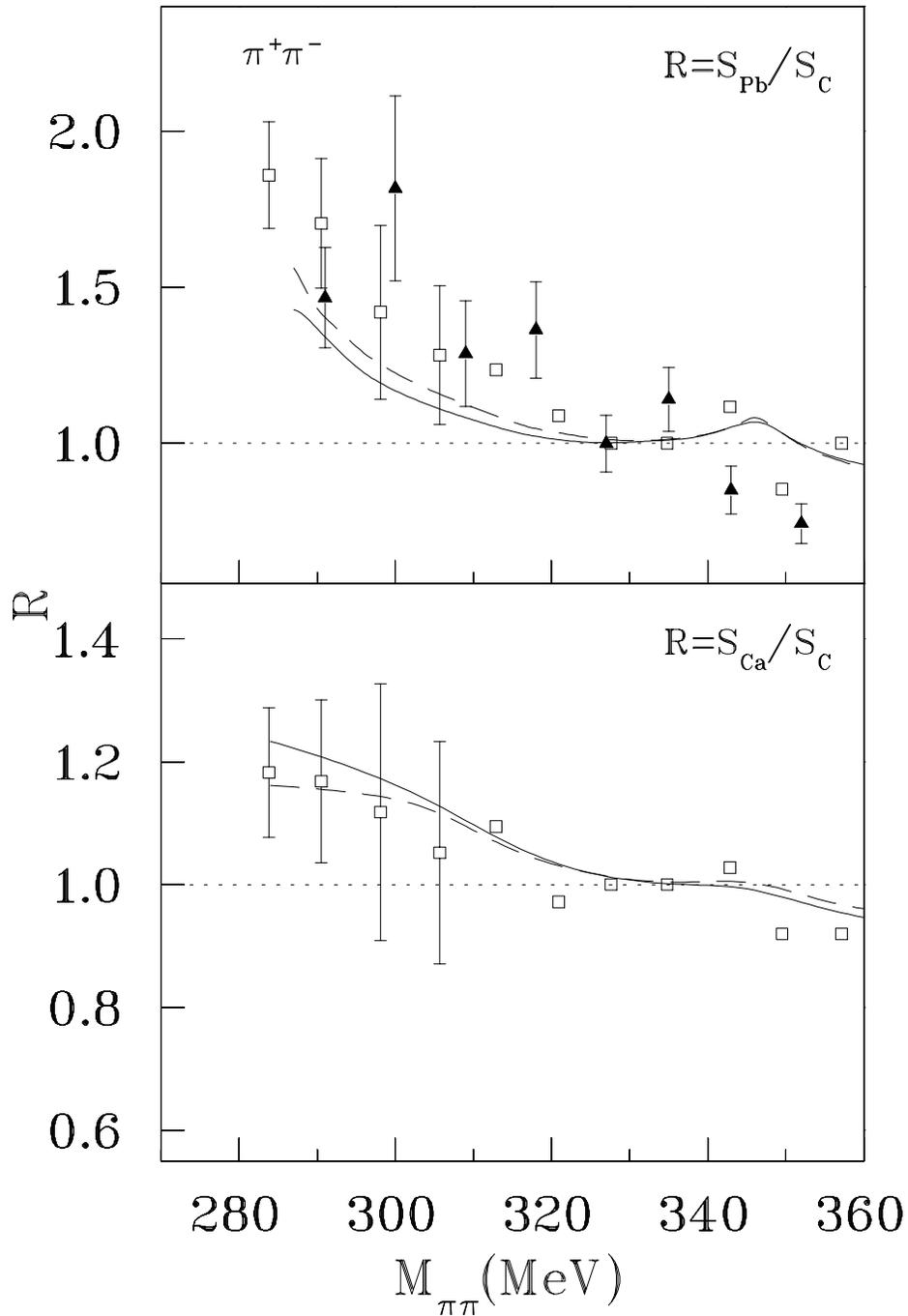}
\caption{Ratios for the $\pi^+\pi^-$ final state compared with the CHAOS
data \protect \cite{chaos99} (squares) and \protect \cite{chaos96}
(triangles). The solid line is the result of a calculation with $r=c$
and the dashed line was made with $r=c+a$.}
\label{pifpm}
\end{figure}

\begin{figure}[p]
\epsfysize=180mm
\epsffile{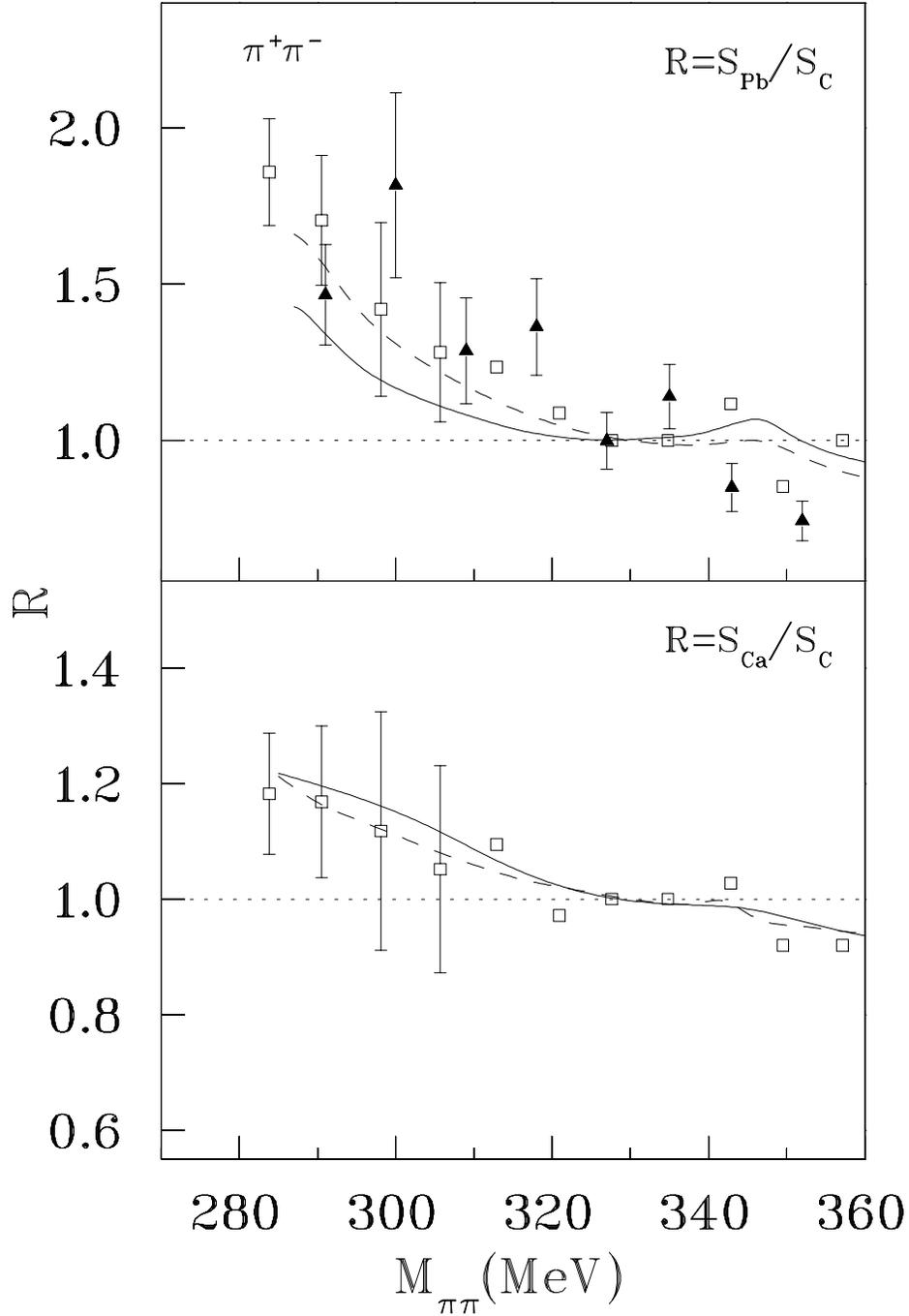}
\caption{The upper panel shows a comparison the ratio calculated
using values of $\lambda_0$ obtained from the fit to gold (solid)
and the light nuclei (dashed). The lower panel compares the result using
the Woods-Saxon densities given in the text (solid line) with the result
of using a shell model density for Ca (dashed line). The data is the same
as in figure \protect \ref{pifpm}.}
\label{lamden}
\end{figure}

\begin{figure}[p]
\epsfysize=180mm
\epsffile{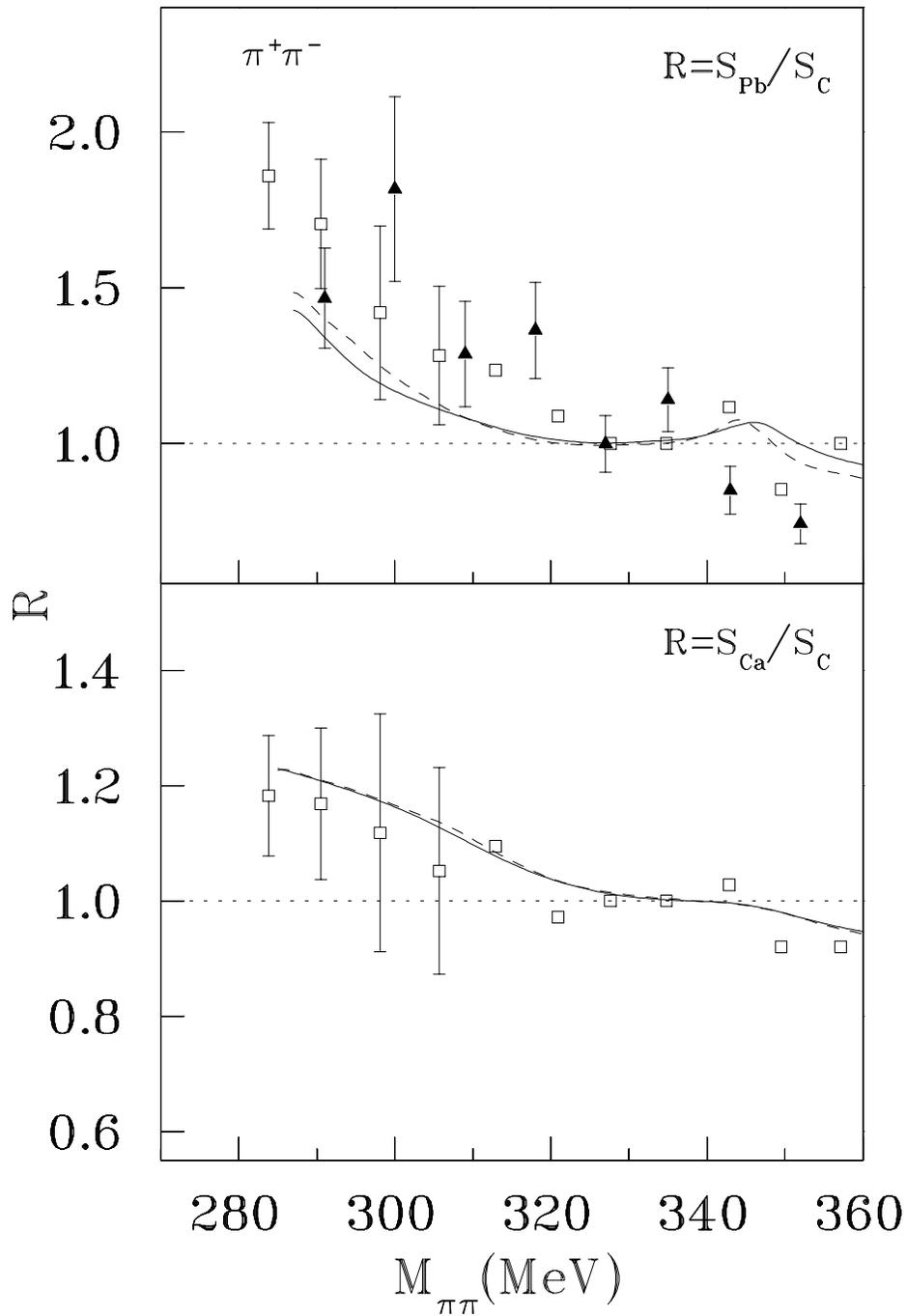}
\caption{Results showing the effect of setting the strong potential to
zero for the $\pi^+\pi^-$ final state. The solid curve is calculated with
the strong potential and the dashed curve has it set to zero.}
\label{pifvs0}
\end{figure}

Having said that, one must admit that there is some variation in the in
the expected density going from (say) carbon to lead.  If we use the
linear dependence of the condensate on the density of Cohen, Furnstahl and
Griegel \cite{cohen} with the calculation of the change in the shape of
the spectrum by Hatsuda and collaborators \cite{hatsuda} along with a
guess of the change in nuclear density, the result is of the same order as
the discrepancy seen in the upper part of Figs. \ref{pifpm} or
\ref{pifvs0}.  However, one must be very careful about trying to draw any
conclusion from this since this also about the same order as the
difference between the two choices in $\lambda_0$ as seen in Fig.
\ref{lamden} and not far outside the experimental errors.

Of course, one might try to compare the production on the nucleon
with production in the nucleus which is close to what has been done 
before.  One would have to assume that the mechanism was simply production 
from one of the nucleons in the nucleus (a reasonable first assumption).  
In the present experiments the comparison has been made with the deuteron.  

It has long been known (see Ref. \cite{kirz} for example) that the
shape of the invariant mass spectrum for the $\pi^+\pi^-$ final pair
is different from that of the $\pi^+\pi^+$ pair at low incident energy.
According to Ref. \cite{oset} the sharp decrease in the low mass region is
due to a cancellation in diagrammatic contributions. These authors suggest
that an alteration of the cancellation due to interaction with the
surrounding nucleons might lead to a modification in this cancellation and
thus to a change in shape of the production spectrum in the direction
seen.

It is also clear (see again Ref. \cite{kirz}) that the $\pi^+\pi^-$
spectrum shape changes to more resemble the $\pi^+\pi^+$ distribution
(and phase space) as the energy increases.  As one goes from the reaction
on a single nucleon to that on a nucleus the effective energy at which the
reaction takes place becomes higher due to the possibilities of coherent
production.  Thus, one should also expect an increase of strength in the 
low mass part of the spectrum from this effect as well. 

This work was supported by the National Science Foundation (Grant 0099729) 
and the Jordan University of Science and Technology.

\newpage

\end{document}